\documentclass[a4paper,11pt]{article}
\usepackage{jheppub}
\usepackage{amsmath,amssymb,amsfonts,graphicx,multirow,color,algorithmic,algorithm}
\usepackage[utf8]{inputenc}
\usepackage{amsmath}
\usepackage{caption}
\usepackage{subcaption}
\usepackage{hyperref}
\usepackage{cleveref}
\usepackage{breakurl} %justify references by breaking urls
\RequirePackage{makecell}
\RequirePackage{xcolor}
\RequirePackage{colortbl}
\graphicspath{{graphics/}}
\makeatletter
\ifx\input@path\@undefined
\def\input@path{{graphics/}}
\else
\g@addto@macro\input@path{{graphics/}}
\fi
\makeatother

\renewcommand{\>}{\rangle}
\newcommand{\<}{\langle}
\newcommand{\Nc}{N_{\mathrm{c}}}

\newcommand{\CVolver}{\texttt{CVolver}}

\preprint{}

\title{Fully Differential Soft Gluon Evolution at the Amplitude Level}

\author[a]{Jeffrey R. Forshaw,}
\author[b,c]{Simon Pl\"atzer,}
\author[a]{and Fernando Torre Gonz\'alez}
\affiliation[a]{Department of Physics and Astronomy, University of
  Manchester, Manchester M13 9PL, United Kingdom}
\affiliation[b]{Institute of Physics,
  NAWI Graz, University of Graz, Universit\"atsplatz 5, A-8010 Graz,
  Austria}
\affiliation[c]{Particle Physics, Faculty of Physics, University of
  Vienna, Boltzmanngasse 5, A-1090 Wien, Austria}

\emailAdd{jeffrey.forshaw@manchester.ac.uk}
\emailAdd{simon.plaetzer@uni-graz.at}
\emailAdd{fernando.torre@manchester.ac.uk}

\date{\today}

\abstract{ We study differential intra-jet radiation patterns in jet
  production at full colour. We present a systematic study of several
  QCD $2\to 2$ processes and also multi-jet production from a colourless
  initial state. We examine how subleading colour
  corrections are distributed differentially in phase space and find
  that mere normalization effects due to subleading colour can be due to subtle cancellations across
  phase space. In general, we find that subleading colour does affect the shapes of distributions.}

\begin{document}

\maketitle

%% start contents %%%%%%%%%%%%%%%%%%%%%%%%%%%%%%%%%%%%%%%%%%%%%%%%%%%%%%%%%%%%%%
%%%%%%%%%%%%%%%%%%%%%%%%%%%%%%%%%%%%%%%%%%%%%%%%%%%%%%%%%%%%%%%%%%%%%%%%%%%%%%%%

\section{Introduction}
\label{sec:Introduction}

In a partner paper \cite{Forshaw:2025bmo}, we presented a systematic
and comprehensive analysis of sub-leading colour corrections in
perturbative QCD processes involving multiple soft gluon emissions. In
that paper, we used a dedicated module of the \CVolver\ code
\cite{Platzer:2013fha,DeAngelis:2020rvq} to study jet veto cross
sections for a range of partonic scattering processes. In this work,
we extend the scope of the analysis by employing \CVolver\ as a
partonic event generator in order to study a broader class of jet
observables, enabling a comprehensive investigation of colour
evolution effects across a variety of final-state
configurations. Specifically, we consider observables fully
differential in the kinematics of the highest energy gluon emitted
outside of the primary jets. This allows us to explore how sub-leading
colour effects are distributed in phase space and to perform a first
investigation into what observables (other than veto cross sections)
may pick up substantial corrections. This work is an important step
on the way to a complete, multi-purpose event generator operating at full colour
accuracy.

The remainder of this paper is structured as follows. In the next
section we introduce the observables that will be the focus of the
rest of the paper. For the case of the jet veto cross section, we show
that the event generator mode delivers identical results to those
reported in the partner paper, obtained using the dedicated mode. This
constitutes a highly non-trivial check of the event generator mode. In
Section~\ref{sec:ObservablesLEP} we consider hard scattering processes
with no incoming coloured particles, such as occur in $e^+e^-$
collisions. We consider four-parton final states, e.g. as in $W^+W^-$
or $ZZ$ decay, and are able to account for colour interference and
reconnection effects. In Section~\ref{sec:ObservablesHadronic} we
consider a range of two-to-two hard scattering processes in a variety
of kinematic configurations. Section~\ref{sec:Conclusions} forms our
conclusions. 

For details of the theoretical framework underpinning this analysis we
refer to the partner paper \cite{Forshaw:2025bmo} and also references
\cite{Martinez:2018ffw,DeAngelis:2020rvq,Platzer:2022jny}. The key
point to appreciate is that \CVolver\ is able to resum leading soft
gluon emissions to a prescribed accuracy in $1/\Nc$ by evolving at the
amplitude level.

\section{Fully differential soft gluon evolution}
\label{sec:Specific}

We use a soft gluon evolution equation with an infrared cutoff in both
energy and collinearity. Presently, we only consider observables that
are completely collinear safe and inclusive in soft gluon emissions
below a characteristic scale, i.e. those observables that are only
sensitive to wide-angle soft gluon emissions. Observables that involve
logarithmic sensitivity to the angular cutoff need to be evolved with
a collinear anomalous dimension
\cite{Forshaw:2019ver,Platzer:2022jny}, a study which we defer to a
future publication in which also hard collinear contributions will be
addressed. Observables sensitive to wide-angle soft gluon emissions
include jet veto cross sections and, more generally, the
triple-differential cross section:
\begin{equation}
  \frac{{\rm d}^3\sigma}{{\rm d}\Omega {\rm d}\rho} =
  \sum_n \int {\rm d}\sigma_n \; 
  \delta\left(\rho - E_i\right)
  \delta\left( \Omega - \Omega_{i}\right) \ ,
\end{equation}
where $i$ labels the soft gluon with the highest energy ($\rho$) that
is emitted in what we refer to as the veto region, i.e. outside
of the regions we identify with jets. Our choice of the veto region
will be process dependent but is always such that it avoids the
regions collinear to the hard partons. A number of interesting
observables can be derived from this differential cross section. In
particular we consider:
\begin{align}
  \frac{{\rm d}^2\Sigma(r)}{{\rm d}\Omega} &= 
    \int_0^{r}{\rm d}\rho   \frac{{\rm d}^3\sigma}{{\rm d}\Omega {\rm d}\rho} \label{eq:obs1}\\
    \frac{{\rm d}\Sigma(r)}{{\rm d}\cos{\theta}} &= 
    \int_0^{2\pi}\int_0^{r}{\rm d}\rho \ {\rm d}\phi  \frac{{\rm d}^3\sigma}{{\rm d}\Omega {\rm d}\rho} \label{eq:obs2}\\
    \frac{{\rm d}\Sigma(r)}{{\rm d}\phi} &= 
    \int_{-1}^{1}\int_0^{r}{\rm d}\rho \ {\rm d}(\cos{\theta})  \frac{{\rm d}^3\sigma}{{\rm d}\Omega {\rm d}\rho} \label{eq:obs3} \ .
\end{align}
We do not perform a normalization of the resulting cross section to
the full colour inclusive cross section of the hard process, allowing
subleading colour contributions to change the normalization as we include
higher orders in $1/\Nc$.  Integrating
Eq.~\eqref{eq:obs1} over the veto region gives the `gaps-between-jets'
cross section with veto energy $r$. One key test of the event
generator mode will be to confirm that doing this integral reproduces
the results in \cite{DeAngelis:2020rvq,Forshaw:2025bmo}. Note that
since we do not include hard-collinear physics the inclusion of
Coulomb (Glauber) gluon exchanges will lead to super-leading
logarithms regulated by the collinear cutoff. As a result, we turn off
Coulomb exchanges for the present paper and will return to perform a
more comprehensive study including their effects in the future. That said, in
Appendix~\ref{app:coulomb}, we do show some results for the jet veto cross section. We intend to explore the effects of Coulomb exchanges more fully in an upcoming study.

We use \CVolver\ to generate events with gluon energies greater than
$\mu = 0.1$ (in units of the maximum gluon energy) and with a
collinear cutoff as described in \cite{Forshaw:2025bmo}. The collinear
cutoff is always chosen to be sufficiently small as to guarantee our
results are independent of it. For the hard scatter process, we
consider a variety of partonic processes each with fixed
kinematics. Apart from the energy and collinear cutoffs, there are no
other generator-level cuts, which means we are fully differential in
the final state phase space, though for this paper we will focus on
the observables defined in equations \eqref{eq:obs1}--\eqref{eq:obs3}.
 
      \begin{figure}
      \centering
        \captionsetup[subfigure]{oneside,margin={0.55cm,0cm}}
        \begin{subfigure}{0.49\linewidth}
            \centering
            \includegraphics[width=1.0\textwidth]{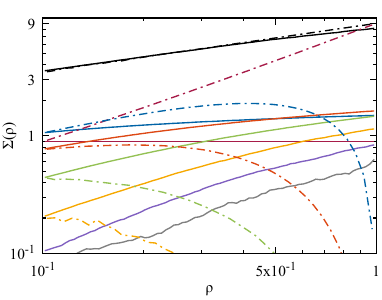}
        \caption{The $|01\>\<01|$ contribution.}
        \label{fig:subfig_a}
        \end{subfigure}
    \begin{subfigure}{0.49\linewidth}
            \centering
            \includegraphics[width=1.0\textwidth]{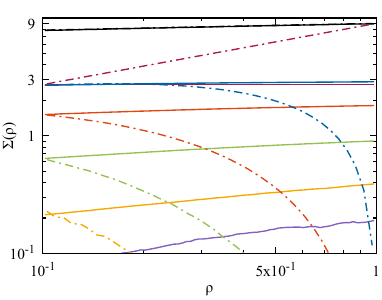}
        \caption{The $|10\>\<10|$ contribution.}
        \label{fig:subfig_b}
        \end{subfigure} \\
     \begin{subfigure}{0.49\linewidth}
            \centering
            \includegraphics[width=1.0\textwidth]{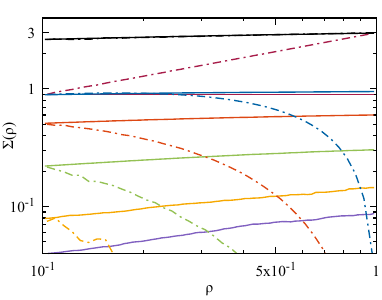}
        \caption{The $|10\>\<01|$ (interference) contribution.}
        \label{fig:subfig_a}
        \end{subfigure}
        \raisebox{0cm}{
     \begin{subfigure}{0.49\linewidth}
            \centering
            \includegraphics[width=0.6\textwidth]{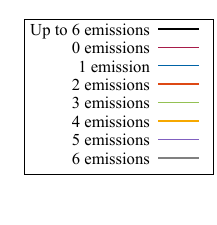}
        \label{fig:legend}
        \end{subfigure}}
     \caption{Comparison of the event generator and dedicated modes of
       \CVolver, for the $q\bar{q} \to q\bar{q}$ process with
       kinematics and colour flows explained in the text. The solid
       curves are the full colour event generator evolution and the
       dash-dotted curves are the full colour dedicated mode
       evolution.}
     \label{fig:hatta-SigmaRhoWithGap}
    \end{figure}
 
An important and highly non-trivial point to check is that we are able
to reproduce our previous results obtained using the dedicated
mode. This is indeed possible and in
Fig.~\ref{fig:hatta-SigmaRhoWithGap} we show the jet veto cross
section, at full colour, for the process $q\bar{q} \to q\bar{q}$ in
the case of back-to-back jets produced at $\pi/6$ to the collision
axis in the zero momentum frame and with the veto region defined by
the central region from $\pi/4 < \theta < 3\pi/4$. The colours used to
represent each multiplicity will remain the same for the rest of this
paper. We use the same notation as in the partner paper to label the
colour-flow density matrix elements of the hard scatter process, which
we dress with soft gluon evolution. Specifically, $|01\>$ is the
colour flow corresponding to $s$-channel singlet exchange and it is
also the leading colour flow corresponding to $t$-channel gluon
exchange, whereas $|10\>$ is the flow corresponding to $t$-channel
singlet exchange and it is also the leading flow corresponding to
$s$-channel gluon exchange.
  
 For the $|10\>\<10|$ and $|10\>\<01|$ colour flow contributions to
 the hard scattering, we observe perfect agreement between the two
 modes (the solid and dash-dotted black curves coincide). This is
 especially non-trivial as the multiplicities build the cross section
 differently: in dedicated mode all emissions are out of gap, while in
 event generator mode emissions can be in any region. The event generator data was evolved down to an infrared scale
 $\mu = 0.1$, and to contribute to the cross section at $\rho=0.1$ all
 emissions must have been out of the gap. Therefore, for the lowest
 $\rho$ bin, the event generator mode and dedicated mode are
 equivalent. The reason for the disagreement at large $\rho$ in the
 $|01\>\<01|$ case is due to missing contributions from higher
 multiplicities since we did not include $\ge 7$ gluon emissions
 and this colour flow radiates more into the gap region. To avoid the
 need to consider more than 6 emissions, in what follows we always
 take $r=0.3$, i.e. we always veto hard radiation emitted in the veto
 region. Note that there is no impediment to 
 reducing $\mu$ and exploring lower values of $\rho$ but we choose not to do
 so because $\rho > 0.1$ is sufficient to see clearly the impact of
 subleading colour.

\section{Final state jets}
\label{sec:ObservablesLEP}

   	\begin{figure}
        \centering
        \captionsetup[subfigure]{oneside,margin={-0.3cm,0cm}}
        \begin{subfigure}{0.49\linewidth}
            \centering
            \includegraphics[width=1.0\textwidth]{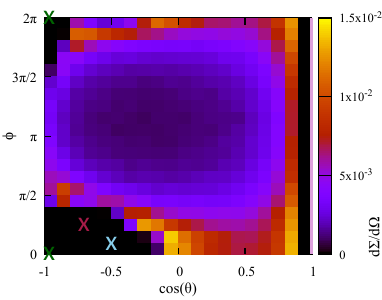}
        \caption{The $|01\>\<01|$ contribution.}
        \label{fig:subfig_a}
        \end{subfigure}
    \begin{subfigure}{0.49\linewidth}
            \centering
            \includegraphics[width=1.0\textwidth]{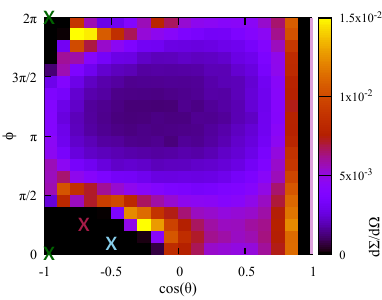}
        \caption{The $|10\>\<10|$ contribution.}
        \label{fig:subfig_b}
        \end{subfigure} \\
     \begin{subfigure}{0.49\linewidth}
            \centering
            \includegraphics[width=1.0\textwidth]{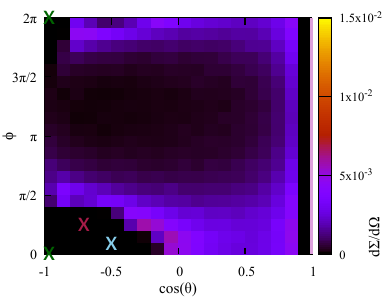}
        \caption{The $|10\>\<01|$ (interference) contribution.}
        \label{fig:subfig_a}
        \end{subfigure}
     \caption{The full colour, three-gluon-emission differential cross section contribution to $\mathrm{d}\Sigma /\mathrm{d} \Omega$ for the $ZZ \to q\bar{q}q\bar{q}$ process. The locations of the hard jets are marked: the four crosses and the pink line. The black regions around the primary partons indicate the jet regions.}
     \label{fig:fullyAsymmetric-2d-FC}
    \end{figure}

In this section our focus is a $q\bar{q}q\bar{q}$ final state, as
might occur in $ZZ \to $ jets. We fixed the kinematics in an
asymmetric fashion and Fig.~\ref{fig:fullyAsymmetric-2d-FC} shows the
differential distribution of the highest energy soft gluon emitted
into the veto region in events with exactly three emissions, i.e. the
$n=3$ contribution to Eq.~\eqref{eq:obs1}. The veto region is defined
as the region outside of the four cones, each with opening angle
$\arccos{(0.95)}$, centred around the hard partons. We did not
generate any more statistics for this process than the other processes
we consider in this paper, so these plots are indicative that we are
able to analyse full colour, soft radiation fully differentially
without much trouble. The locations of the four hard partons are
marked on the plot. The vertical pink line on the right corresponds to a quark emitted at
$\theta=0$, the green crosses are a quark emitted at $\theta =
11\pi/12$ and $\phi=0$, the light blue cross is an anti-quark at
$\theta = 2\pi/3$ and $\phi=\pi/12$ and the red cross is an anti-quark
at $\theta = 3\pi/4$ and $\phi=\pi/4$. The colour-flow configurations
are such that $|01\>$ corresponds to the pink (quark) and light blue
(anti-quark) particles being colour connected (and also the green and
red particles) whereas the $|10\>$ flow corresponds to the pink and
red being connected (and the blue and green).  The black regions
surrounding the hard partons are the jet regions (over which we are fully inclusive), i.e. the regions
complementary to the veto region. A noticeable feature is the strong
radiation around $\cos \theta \sim 0.4$ between pink quark and light blue anti-quark in the
$|01\>\<01|$ plot, which is much weaker for the $|10\>\<10|$
contribution. In the latter, the pink quark is colour connected to the
red anti-quark, which shifts the radiation higher in $\phi$. The
radiation of the $|10\>\<01|$ interference contribution is an entirely
subleading colour effect, because the interference density matrix
cannot emit at leading colour. This quantity is of particular
interest, since electroweak diboson final states at LEP2 have been considered as key tools to investigate colour
reconnection models \cite{L3:2003oci,DELPHI:2006tie}, and the $W$ mass
extraction at future lepton colliders will rely on how
well colour reconnection effects between the hadronic $W$ systems
are under control \cite{deBlas:2024bmz}.

 \begin{figure}
        \centering
        \captionsetup[subfigure]{oneside,margin={0.97cm,0cm}}
        \begin{subfigure}{0.49\linewidth}
            \centering
            \includegraphics[width=1.0\textwidth]{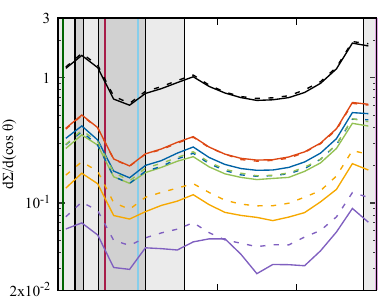}
        \end{subfigure}
    \begin{subfigure}{0.49\linewidth}
            \centering
            \includegraphics[width=1.0\textwidth]{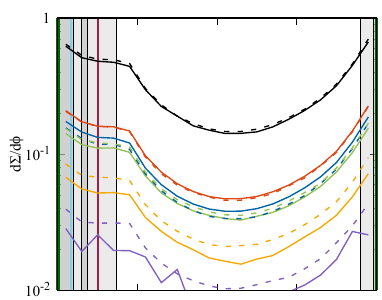}
        \end{subfigure} \\
        \begin{subfigure}{0.49\linewidth}
            \centering
            \includegraphics[width=1.0\textwidth]{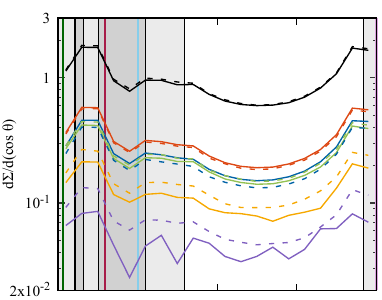}
        \end{subfigure}
    \begin{subfigure}{0.49\linewidth}
            \centering
            \includegraphics[width=1.0\textwidth]{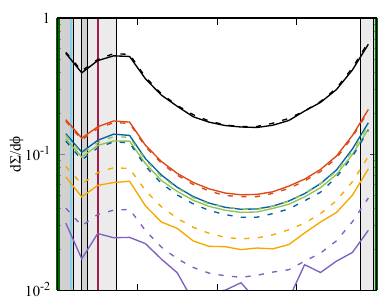}
        \end{subfigure} \\
        \begin{subfigure}{0.49\linewidth}
            \centering
            \includegraphics[width=1.0\textwidth]{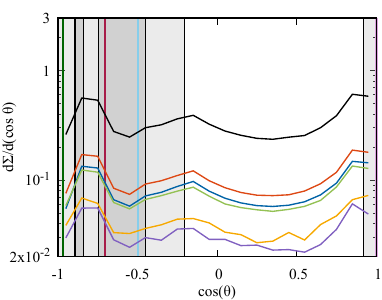}
        \end{subfigure}
    \begin{subfigure}{0.49\linewidth}
            \centering
            \includegraphics[width=1.0\textwidth]{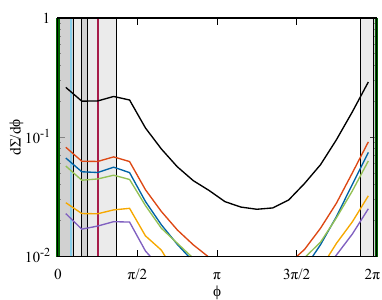}
        \end{subfigure} \\
     \caption{The differential cross sections for the different
       contributions to the $ZZ \to q\bar{q}q\bar{q}$ process, broken
       down by multiplicity. The top row is the $|01\>\<01|$
       contribution, the middle row is the $|10\>\<10|$ contribution,
       and the bottom row is the $|10\>\<01|$ (interference)
       contribution. The solid curves are the full colour evolution,
       and the dashed curves are the leading colour evolution. For the
       interference case there is no leading colour contribution
       because there are no dipoles to emit from. The locations of the
       hard jets are marked with vertical lines matching the colours
       used in Fig.~\ref{fig:fullyAsymmetric-2d-FC}. The shaded
       vertical bars indicate the width of the cones around each hard
       parton. Darker shades indicate an overlap of multiple cones.}
     \label{fig:fullyAsymmetric-interferences}
    \end{figure}
  
In Fig.~\ref{fig:fullyAsymmetric-interferences} we show the cross
section differential in the polar and azimuthal angles of the highest
energy soft gluon emitted into the veto region,
i.e. Eq.~\eqref{eq:obs2} and Eq.~\eqref{eq:obs3}. We find that there
is good agreement between leading and full colour results for the
$|01\>\<01|$ and $|10\>\<10|$ contributions. However, the interference
distributions have significantly different radiation patterns compared
to the colour-diagonal contributions. In particular, the $\mathrm{d}
\Sigma / \mathrm{d} \phi$ interference contribution is much steeper in
the region away from the hard jets.

\section{Hadronic jets}
\label{sec:ObservablesHadronic}

    \begin{figure}
      \centering
      \includegraphics[width=0.7\textwidth]{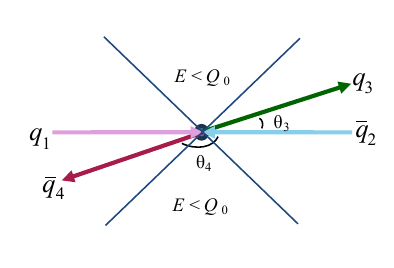}
     \caption{The kinematic configuration of the four hard partons in
       $q_1 \bar{q}_2 \to q_3 \bar{q}_4$. The angles of the different
       configurations considered are given in
       Table~\ref{tab:kinematics}, and the veto region definitions for
       each configuration are given in the text. The colours of the
       four legs (pink $q_1$, light blue $\bar{q}_2$, dark green
       $q_3$, red $\bar{q}_4$) are used in the figures to indicate the
       location of the hard partons. The colour state $|01\>$
       corresponds to pink-light blue colour connected, and dark
       green-red connected. The colour state $|10\>$ corresponds to
       pink-dark green connected, and light blue-red connected.}
     \label{fig:hatta-diagram}
    \end{figure}

\begin{table}[t]
    \centering
    \begin{tabular}{|c!{\vrule width 1.7pt}c|c!{\vrule width 1.7pt}c|c!{\vrule width 1.7pt}c|c!{\vrule width 1.7pt}}
      \cline{2-7}
      \multicolumn{1}{l!{\vrule width 1.7pt}}{} & \multicolumn{2}{c!{\vrule width 1.7pt}}{\makecell{back-to-back \\ }} & \multicolumn{2}{c!{\vrule width 1.7pt}}{\makecell{boosted \\ }} & \multicolumn{2}{c!{\vrule width 1.7pt}}{\makecell{recoiling \\ }} \\
      \cline{2-7}
      \multicolumn{1}{l!{\vrule width 1.7pt}}{} & $\theta$ & $\phi$ & $\theta$ & $\phi$ & $\theta$ & $\phi$ \\
      \hline
   $q_1$ & $0$ & $0$ & $0$ & $0$ & $0$ & $0$ \\
        \arrayrulecolor{gray!70} \hline \arrayrulecolor{black}
   $\bar{q}_2$ & $\pi$ & $0$ & $\pi$ & $0$ & $\pi$ & $0$ \\
        \arrayrulecolor{gray!70} \hline \arrayrulecolor{black}
   $q_3$ & $\pi/6$ & $0$ & $0.037$ & $0$ & $\pi/6$ & $0$ \\
        \arrayrulecolor{gray!70} \hline \arrayrulecolor{black}
   $\bar{q}_4$ & $5\pi/6$ & $\pi$ & $2.873$ & $\pi$ & $\pi/3$ & $19\pi/12$ \\
        \hline
    \end{tabular}
    \caption{The directions of the quarks and anti-quarks for each of the kinematic configurations considered. }
    \label{tab:kinematics}
\end{table}

In this section our focus is the hard scattering $q\bar{q} \to
q\bar{q}$, such as might occur in the collision of two hadrons. We consider three different kinematic configurations, as
specified in Table~\ref{tab:kinematics}. The back-to-back
configuration corresponds to parton scattering in the zero momentum
frame to produce jets in the forward and backward directions. In this
case, the veto region is the central region satisfying $-0.8 < \cos
\theta < 0.8$. The boosted configuration corresponds to a more extreme
low-angle scattering with a boost along the collision axis. In this
case, the veto region is the central region satisfying $-0.9 < \cos
\theta < 0.9$. The recoiling configuration corresponds to the case
where the outgoing partons are not back-to-back in azimuth, such as
would occur if the the jets are recoiling against a colourless
particle (e.g. as in the case of weak vector boson fusion at the
LHC). In this case, the veto region is defined to be the region
outside of jets defined by cones centred on the final state quark and
anti-quark, each with opening angle $\arccos(0.95)$.

In what follows, we will show results arising from the $|01\>\<01|$
and $|10\>\<10|$ hard process density matrices, and also the
combinations corresponding to $t$-channel and $s$-channel gluon
exchange. The $|01\>\<01|$ colour configuration corresponds to the
incoming particles being colour connected and the outgoing particles
being colour connected. The $|10\>\<10|$ configuration corresponds to
the incoming quark connected to the outgoing quark and equivalently
for the antiquarks.
  
  For each contribution we show the result of full colour evolution
  compared to the result of strictly leading colour evolution. For the
  $s$- and $t$- channel gluon exchange contributions, we initiate the
  leading colour evolution using the leading-colour part of the
  hard-scatter matrix. This is what we referred to as the L1, LCH
  approximation in the partner paper and we have chosen it here since
  that is generally the best performing approximation to the full
  colour result for the gaps-between-jets observable: specifically, we
  found that L1, LCH showed very good agreement with the full colour
  result for $t$-channel gluon exchange in every $2 \to 2$ scattering
  process we considered (as first observed in
  \cite{Hatta:2013iba,Hatta:2020wre}). By analysing the soft radiation
  differentially, we aim to stress-test the success of this remarkable
  approximation.
  
\subsection{Back-to-back jets}
\label{sec:b2b}

    \begin{figure}
        \centering
        \captionsetup[subfigure]{oneside,margin={0.97cm,0cm}}
        \begin{subfigure}{0.49\linewidth}
            \begin{subfigure}[t]{1.0\textwidth}
            \centering
            \includegraphics[width=1.0\textwidth]{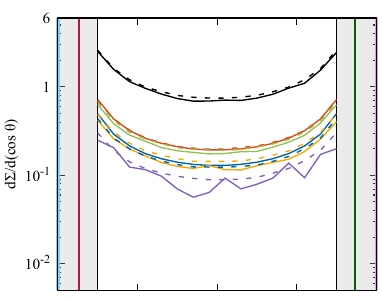}
            \end{subfigure} \\
            \begin{subfigure}[t]{01.\textwidth}
            \centering
            \includegraphics[width=1.0\textwidth]{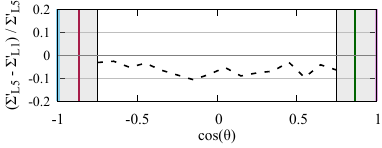}
        \end{subfigure} \\
        \caption{The $|01\>\<01|$ contribution.}
        \label{fig:subfig_a}
    \end{subfigure}
    \begin{subfigure}{0.49\linewidth}
            \begin{subfigure}[t]{1.\textwidth}
            \centering
            \includegraphics[width=1.0\textwidth]{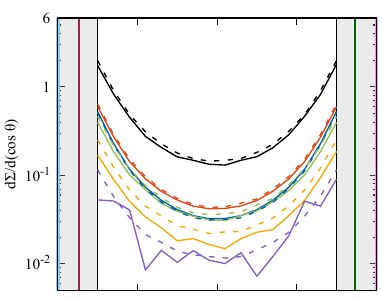}
            \end{subfigure} \\
            \begin{subfigure}[t]{1.\textwidth}
            \centering
            \includegraphics[width=1.0\textwidth]{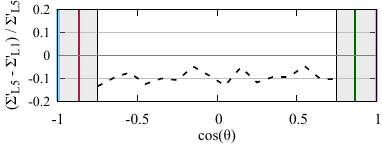}
        \end{subfigure} \\
        \caption{The $|10\>\<10|$ contribution.}
        \label{fig:subfig_b}
     \end{subfigure}
        \captionsetup[subfigure]{oneside,margin={0.02cm,0cm}}
    \begin{subfigure}{0.49\linewidth}
            \begin{subfigure}[t]{1.\textwidth}
            \centering
            \includegraphics[width=1.0\textwidth]{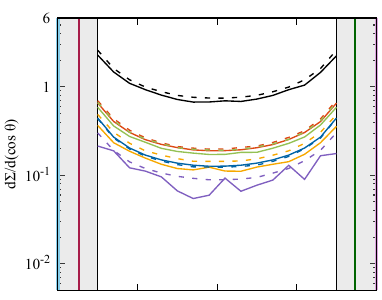}
            \end{subfigure} \\
            \begin{subfigure}[t]{1.\textwidth}
            \centering
            \includegraphics[width=1.0\textwidth]{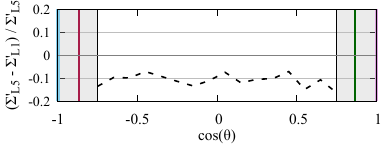}
        \end{subfigure} \\
        \caption{The $t$-channel gluon exchange contribution.}
        \label{fig:subfig_b}
     \end{subfigure}
     \begin{subfigure}{0.49\linewidth}
            \begin{subfigure}[t]{1.0\textwidth}
            \centering
            \includegraphics[width=1.0\textwidth]{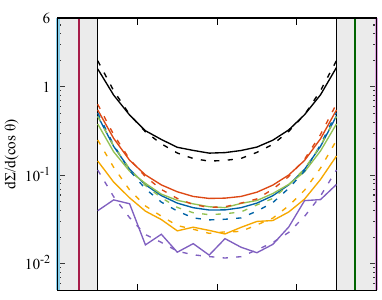}
            \end{subfigure} \\
            \begin{subfigure}[t]{01.\textwidth}
            \centering
            \includegraphics[width=1.0\textwidth]{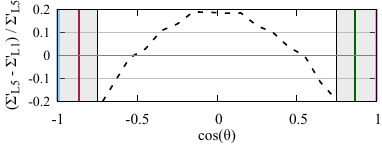}
        \end{subfigure} \\
        \caption{The $s$-channel gluon exchange contribution.}
        \label{fig:subfig_a}
    \end{subfigure}
     \caption{ The differential cross section $\mathrm{d}\Sigma /
       \mathrm{d}(\cos{\theta})$ of the different contributions to the
       $q\bar{q} \to q\bar{q}$ process, in the back-to-back
       configuration, and broken down by multiplicity. The solid
       curves are obtained using full colour evolution, and the dashed
       curves use strictly leading colour evolution. For the leading
       colour curves, we start the evolution using the leading-colour
       approximation to the hard-scatter density matrix. The locations
       of the hard jets are marked with vertical lines matching the
       colours used in Fig.~\ref{fig:hatta-diagram}. The shaded
       vertical bars indicate the jet regions.}
     \label{fig:hatta-cosTheta}
    \end{figure}

\begin{figure}
        \centering
        \captionsetup[subfigure]{oneside,margin={0.97cm,0cm}}
        \begin{subfigure}{0.49\linewidth}
            \centering
            \includegraphics[width=1.0\textwidth]{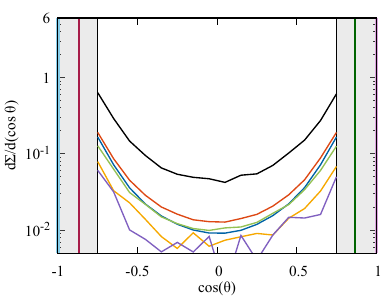}
        \end{subfigure}
    \begin{subfigure}{0.49\linewidth}
            \centering
            \includegraphics[width=1.0\textwidth]{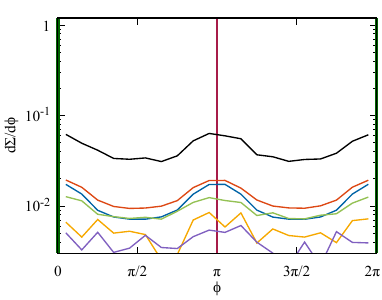}
        \end{subfigure} 
     \caption{ The differential cross sections arising from the
       $|10\>\<01|$ (interference) contribution to the $q\bar{q} \to
       q\bar{q}$ process, in the back-to-back configuration, and
       broken down by multiplicity. The solid curves are obtained
       using full colour evolution. There is no leading colour
       contribution since there are no dipoles to emit from. The
       locations of the hard jets are marked with vertical lines
       matching the colours used in Fig.~\ref{fig:hatta-diagram}. The
       shaded vertical bars indicate the jet regions.}
     \label{fig:hatta-interferences}
    \end{figure}
    
    \begin{figure}
        \centering
        \captionsetup[subfigure]{oneside,margin={0.97cm,0cm}}
        \begin{subfigure}{0.49\linewidth}
            \begin{subfigure}[t]{1.0\textwidth}
            \centering
            \includegraphics[width=1.0\textwidth]{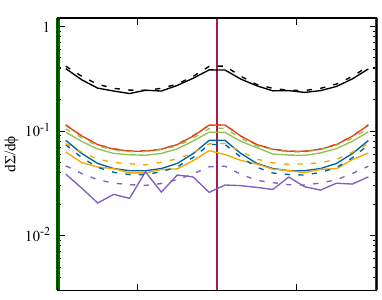}
            \end{subfigure} \\
            \begin{subfigure}[t]{01.\textwidth}
            \centering
            \includegraphics[width=1.0\textwidth]{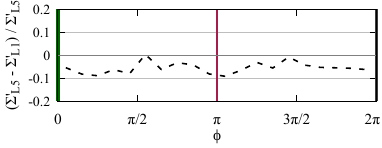}
        \end{subfigure} \\
        \caption{The $|01\>\<01|$ contribution.}
        \label{fig:subfig_a}
    \end{subfigure}
    \begin{subfigure}{0.49\linewidth}
            \begin{subfigure}[t]{1.\textwidth}
            \centering
            \includegraphics[width=1.0\textwidth]{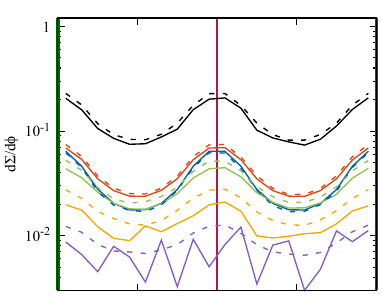}
            \end{subfigure} \\
            \begin{subfigure}[t]{1.\textwidth}
            \centering
            \includegraphics[width=1.0\textwidth]{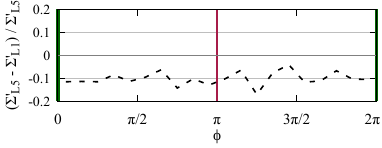}
        \end{subfigure} \\
        \caption{The $|10\>\<10|$ contribution.}
        \label{fig:subfig_b}
     \end{subfigure}
        \captionsetup[subfigure]{oneside,margin={0.02cm,0cm}}
    \begin{subfigure}{0.49\linewidth}
            \begin{subfigure}[t]{1.\textwidth}
            \centering
            \includegraphics[width=1.0\textwidth]{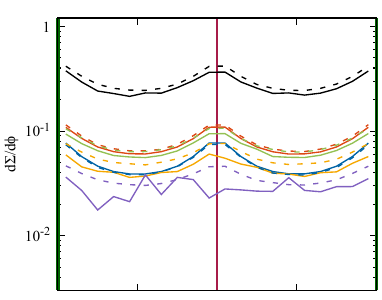}
            \end{subfigure} \\
            \begin{subfigure}[t]{1.\textwidth}
            \centering
            \includegraphics[width=1.0\textwidth]{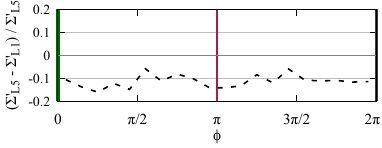}
        \end{subfigure} \\
        \caption{The $t$-channel gluon exchange contribution.}
        \label{fig:subfig_b}
     \end{subfigure}
     \begin{subfigure}{0.49\linewidth}
            \begin{subfigure}[t]{1.0\textwidth}
            \centering
            \includegraphics[width=1.0\textwidth]{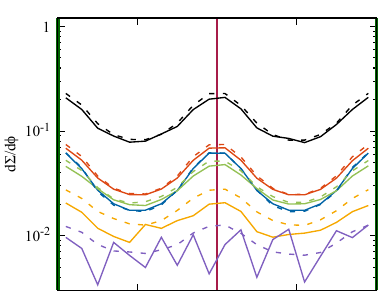}
            \end{subfigure} \\
            \begin{subfigure}[t]{01.\textwidth}
            \centering
            \includegraphics[width=1.0\textwidth]{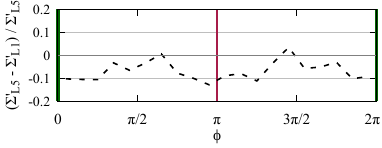}
        \end{subfigure} \\
        \caption{The $s$-channel gluon exchange contribution.}
        \label{fig:subfig_a}
    \end{subfigure}
     \caption{ The differential cross section $\mathrm{d}\Sigma /
       \mathrm{d}\phi$ of the different contributions to the $q\bar{q}
       \to q\bar{q}$ process, in the back-to-back configuration, and
       broken down by multiplicity. The solid curves are obtained
       using full colour evolution, and the dashed curves use strictly
       leading colour evolution. For the leading colour curves, we
       start the evolution using the leading-colour approximation to
       the hard-scatter density matrix. The locations of the hard jets
       are marked with vertical lines matching the colours used in
       Fig.~\ref{fig:hatta-diagram}.}
     \label{fig:hatta-phi}
    \end{figure}

  The $\mathrm{d}\Sigma / \mathrm{d}(\cos{\theta})$ distributions are
  shown in Fig.~\ref{fig:hatta-cosTheta}. As expected, we see the
  radiation grows as it gets closer to the hard jets (their location
  is indicated with vertical lines matching the colours in
  Fig.~\ref{fig:hatta-diagram}). The $|01\>\<01|$ distribution is less
  steep than $|10\>\<10|$, which follows from the colour flows: in the
  former configuration, the colour-connected pairs are back-to-back,
  which causes them to emit in all directions, including the veto
  region. In contrast, $|10\>\<10|$ has the connected pairs at angles
  of $\pi/6$, which leads to most of the radiation being emitted out
  of the veto region, and results in a much steeper radiation pattern.
  
  Since the strictly leading colour and full colour results have
  different normalisations, we should focus on the shape of the
  residuals. We see that the residual for the $|10\>\<10|$
  contribution is approximately flat. In contrast, the residual for
  the $|01\>\<01|$ shows a $5$-$10\%$ effect between the edges and
  centre of the veto region, although we have some fluctuations. It is
  perhaps easier to see this effect by looking at the 1, 2 or 3
  emission curves (blue, orange, light green respectively). The effect
  is caused by virtual gluon exchange causing swaps to the
  $|10\>\<10|$ configuration, which radiates more outside of the veto
  region, therefore enhancing the differential cross section near the
  edges of the veto region relative to the centre. That said, we
  have a clearer example in the next section.
  
  The $t$-channel gluon exchange contribution has a flat residual,
  showing that those subleading-colour effects in the $|01\>\<01|$
  evolution cancel, but only after also including the interference
  contribution, shown in Fig.~\ref{fig:hatta-interferences}, and the
  subleading $|10\>\<10|$ contribution. This is despite the fact that
  the interference evolution vanishes at leading colour. Moreover, the
  radiation pattern from the interference contribution has a very
  different shape (steeper) compared to those of the $|01\>\<01|$ and
  $|10\>\<10|$ contributions.
  
  The $s$-channel contribution shows the largest discrepancy between
  the full colour and leading colour results. The inclusion of the
  subleading $|01\>\<01|$ and $|10\>\<01|$ interference contributions
  flattens the full colour radiation pattern, causing an overall
  $40\%$ effect between the edges and the middle of the veto region.
  
  We show the $\mathrm{d}\Sigma / \mathrm{d} \phi$ distributions in
  Fig.~\ref{fig:hatta-phi}. The positions in $\phi$ of the two
  outgoing particles are shown with vertical lines on the plot. Again,
  we see that the $|01\>\<01|$ contribution radiates more evenly
  in $\phi$ than the $|10\>\<10|$ contribution, which prioritises
  radiation along the directions of the two colour-connected
  pairs. The $|10\>\<10|$ residual is approximately flat, while the
  $|01\>\<01|$ residual shows $\sim 5\%$ effects between the edges and
  centre of the veto region. The $t$-channel residual is approximately
  flat, although with fluctuations, while the $s$-channel residual
  shows $\sim 10\%$ effects.

\subsection{Jets with a boost}

    \begin{figure}
        \centering
        \captionsetup[subfigure]{oneside,margin={0.97cm,0cm}}
        \begin{subfigure}{0.49\linewidth}
            \begin{subfigure}[t]{1.0\textwidth}
            \centering
            \includegraphics[width=1.0\textwidth]{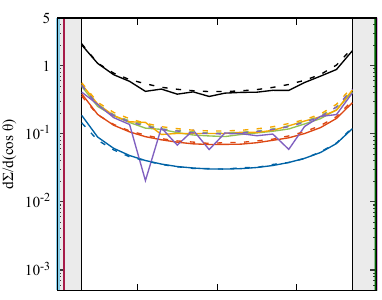}
            \end{subfigure} \\
            \begin{subfigure}[t]{01.\textwidth}
            \centering
            \includegraphics[width=1.0\textwidth]{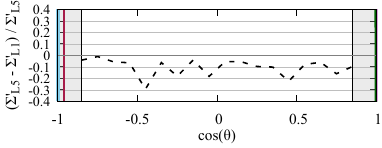}
        \end{subfigure} \\
        \caption{The $|01\>\<01|$ contribution.}
        \label{fig:subfig_a}
    \end{subfigure}
    \begin{subfigure}{0.49\linewidth}
            \begin{subfigure}[t]{1.\textwidth}
            \centering
            \includegraphics[width=1.0\textwidth]{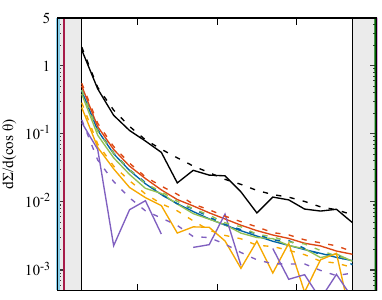}
            \end{subfigure} \\
            \begin{subfigure}[t]{1.\textwidth}
            \centering
            \includegraphics[width=1.0\textwidth]{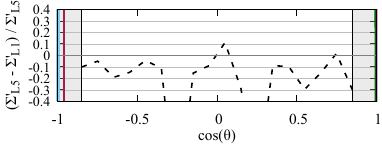}
        \end{subfigure} \\
        \caption{The $|10\>\<10|$ contribution.}
        \label{fig:subfig_b}
     \end{subfigure}
        \captionsetup[subfigure]{oneside,margin={0.02cm,0cm}}
    \begin{subfigure}{0.49\linewidth}
            \begin{subfigure}[t]{1.\textwidth}
            \centering
            \includegraphics[width=1.0\textwidth]{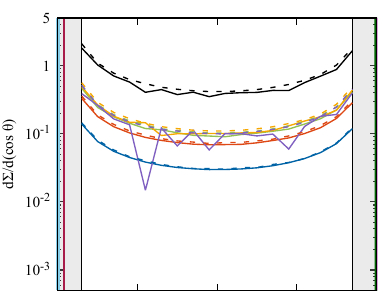}
            \end{subfigure} \\
            \begin{subfigure}[t]{1.\textwidth}
            \centering
            \includegraphics[width=1.0\textwidth]{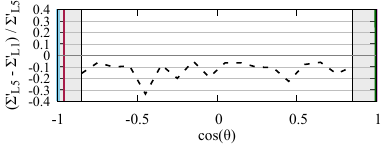}
        \end{subfigure} \\
        \caption{The $t$-channel gluon exchange contribution.}
        \label{fig:subfig_b}
     \end{subfigure}
     \begin{subfigure}{0.49\linewidth}
            \begin{subfigure}[t]{1.0\textwidth}
            \centering
            \includegraphics[width=1.0\textwidth]{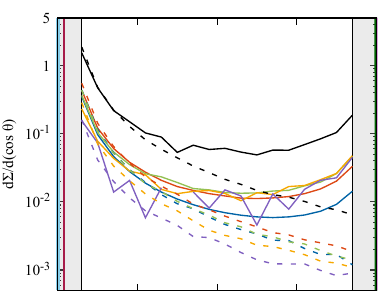}
            \end{subfigure} \\
            \begin{subfigure}[t]{01.\textwidth}
            \centering
            \includegraphics[width=1.0\textwidth]{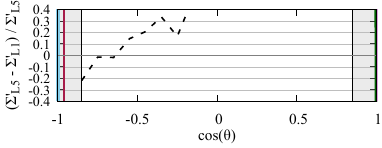}
        \end{subfigure} \\
        \caption{The $s$-channel gluon exchange contribution.}
        \label{fig:subfig_a}
    \end{subfigure}
     \caption{ The differential cross section $\mathrm{d}\Sigma /
       \mathrm{d}(\cos{\theta})$ of the different contributions to the
       $q\bar{q} \to q\bar{q}$ process, in the boosted jets
       configuration, and broken down by multiplicity. The solid
       curves are obtained using full colour evolution, and the dashed
       curves use strictly leading colour evolution. For the leading
       colour curves, we start the evolution using the leading-colour
       approximation to the hard-scatter density matrix. The locations
       of the hard jets are marked with vertical lines matching the
       colours used in Fig.~\ref{fig:hatta-diagram}. The shaded
       vertical bars indicate the jet regions.}
     \label{fig:phiSymmetric-cosTheta}
    \end{figure}

    \begin{figure}
        \centering
        \captionsetup[subfigure]{oneside,margin={0.97cm,0cm}}
        \begin{subfigure}{0.49\linewidth}
            \begin{subfigure}[t]{1.0\textwidth}
            \centering
            \includegraphics[width=1.0\textwidth]{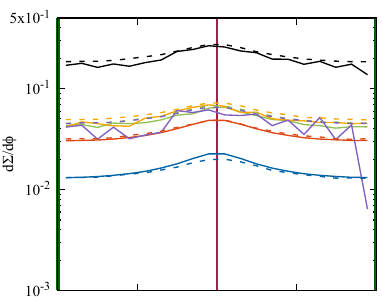}
            \end{subfigure} \\
            \begin{subfigure}[t]{01.\textwidth}
            \centering
            \includegraphics[width=1.0\textwidth]{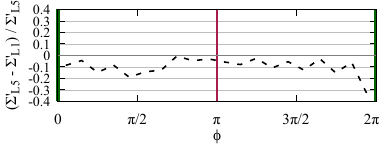}
        \end{subfigure} \\
        \caption{The $|01\>\<01|$ contribution.}
        \label{fig:subfig_a}
    \end{subfigure}
    \begin{subfigure}{0.49\linewidth}
            \begin{subfigure}[t]{1.\textwidth}
            \centering
            \includegraphics[width=1.0\textwidth]{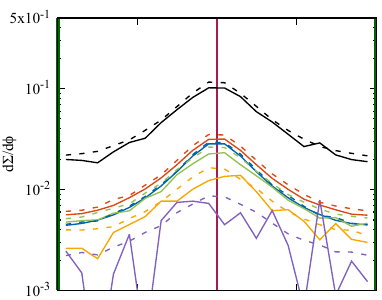}
            \end{subfigure} \\
            \begin{subfigure}[t]{1.\textwidth}
            \centering
            \includegraphics[width=1.0\textwidth]{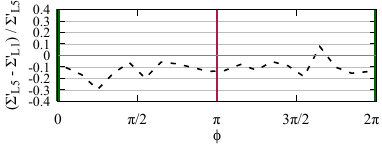}
        \end{subfigure} \\
        \caption{The $|10\>\<10|$ contribution.}
        \label{fig:subfig_b}
     \end{subfigure}
        \captionsetup[subfigure]{oneside,margin={0.02cm,0cm}}
    \begin{subfigure}{0.49\linewidth}
            \begin{subfigure}[t]{1.\textwidth}
            \centering
            \includegraphics[width=1.0\textwidth]{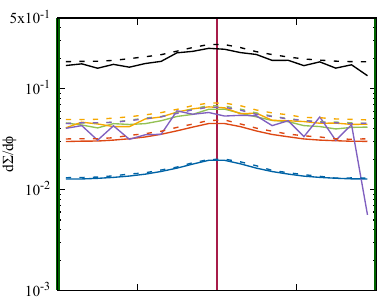}
            \end{subfigure} \\
            \begin{subfigure}[t]{1.\textwidth}
            \centering
            \includegraphics[width=1.0\textwidth]{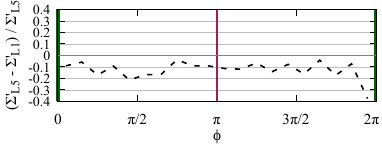}
        \end{subfigure} \\
        \caption{The $t$-channel gluon exchange contribution.}
        \label{fig:subfig_b}
     \end{subfigure}
     \begin{subfigure}{0.49\linewidth}
            \begin{subfigure}[t]{1.0\textwidth}
            \centering
            \includegraphics[width=1.0\textwidth]{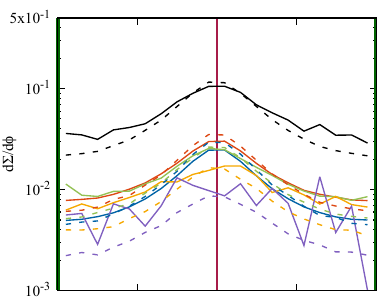}
            \end{subfigure} \\
            \begin{subfigure}[t]{01.\textwidth}
            \centering
            \includegraphics[width=1.0\textwidth]{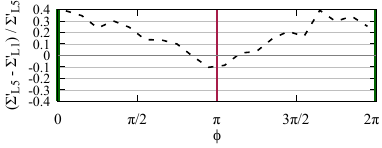}
        \end{subfigure} \\
        \caption{The $s$-channel gluon exchange contribution.}
        \label{fig:subfig_a}
    \end{subfigure}
     \caption{ The differential cross section $\mathrm{d}\Sigma /
       \mathrm{d}\phi$ of the different contributions to the $q\bar{q}
       \to q\bar{q}$ process, in the boosted jets configuration, and
       broken down by multiplicity. The solid curves are obtained
       using full colour evolution, and the dashed curves use strictly
       leading colour evolution. For the leading colour curves, we
       start the evolution using the leading-colour approximation to
       the hard-scatter density matrix. The locations of the hard jets
       are marked with vertical lines matching the colours used in
       Fig.~\ref{fig:hatta-diagram}.}
     \label{fig:phiSymmetric-phi}
    \end{figure}

    \begin{figure}
        \centering
        \captionsetup[subfigure]{oneside,margin={0.97cm,0cm}}
        \begin{subfigure}{0.49\linewidth}
            \centering
            \includegraphics[width=1.0\textwidth]{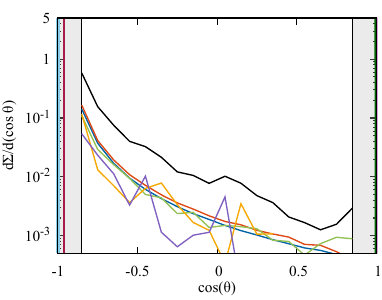}
        \end{subfigure}
    \begin{subfigure}{0.49\linewidth}
            \centering
            \includegraphics[width=1.0\textwidth]{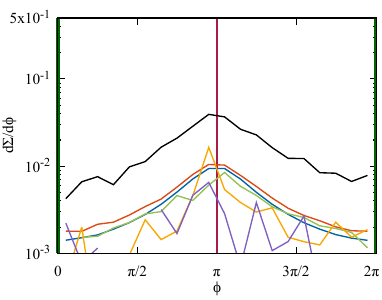}
        \end{subfigure} 
     \caption{ The differential cross sections arising from the
       $|10\>\<01|$ (interference) contribution to the $q\bar{q} \to
       q\bar{q}$ process, in the boosted jets configuration, and
       broken down by multiplicity. The solid curves are obtained
       using full colour evolution. There is no leading colour
       contribution since there are no dipoles to emit from. The
       locations of the hard jets are marked with vertical lines
       matching the colours used in Fig.~\ref{fig:hatta-diagram}. The
       shaded vertical bars indicate the jet regions.}
     \label{fig:phiSymmetric-interferences}
    \end{figure}

We next consider the boosted jets kinematic configuration.  The
$\mathrm{d} \Sigma / \mathrm{d} (\cos{\theta})$ distribution in
Fig.~\ref{fig:phiSymmetric-cosTheta} clearly shows the effect of the
small, and asymmetric, scattering angles: the $|10\>\<10|$
contribution is now highly asymmetric in $\theta$. The $q_1$-$q_3$
opening angle is much smaller than the $\bar{q}_2$-$\bar{q}_4$ opening
angle, suppressing radiation from the former dipole and enhancing it
from the latter. On the other hand, the $|01\>\<01|$ contribution,
where the colour-connected pairs are almost back-to-back, remains more
symmetric, but not completely: full colour evolution can lead to virtual gluon
exchanges that swap from this colour configuration to the other, which
visibly enhances the full colour curves towards
$\cos{\theta}=-1$. This is most visible for the one-emission curve
(solid blue), but is present for all multiplicities.
The $|10\>\<10|$ evolution emits mostly outside the veto region, and
  therefore disfavours swapping to a different colour
  configuration.

  The $s$-channel gluon exchange contribution is very different
  between full and leading colour, as the latter is unable to swap into the
  $|01\rangle \langle 01|$ configuration. Interestingly, the $t$-channel, full colour result once
  again exhibits the cancellation of the small enhancement near
  $\cos{\theta} = -1$ from the $|10\>\<10|$ evolution with the
  interference (Fig.~\ref{fig:phiSymmetric-interferences}) and the
  subleading $|10\>\<10|$ contributions.

  The azimuthal distributions are shown in
  Figs.~\ref{fig:phiSymmetric-phi} and the interference distributions
  are shown in Fig.~\ref{fig:phiSymmetric-interferences}. We note that
  all show the same characteristics: the $|10\>\<10|$ contribution
  shows very good agreement between full and leading colour; the
  $|01\>\<01|$ contribution shows full colour enhances the radiation
  in the region near the hard jets; $s$-channel gluon exchange
  completely fails to be described by leading colour
  evolution; $t$-channel gluon exchange demonstrates that the
  enhancement in $|01\>\<01|$ cancels with the $|10\>\<10|$ and
  interference contributions, resulting in the surprisingly good agreement between full
  colour and leading colour.

\subsection{Jets with a recoil}

\begin{figure}
        \centering
        \captionsetup[subfigure]{oneside,margin={0.02cm,0cm}}
    \begin{subfigure}{0.49\linewidth}
            \begin{subfigure}[t]{1.\textwidth}
            \centering
            \includegraphics[width=1.0\textwidth]{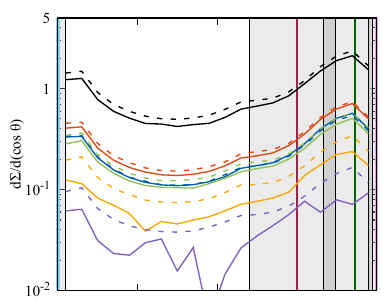}
            \end{subfigure} \\
            \begin{subfigure}[t]{1.\textwidth}
            \centering
            \includegraphics[width=1.0\textwidth]{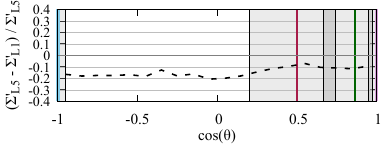}
        \end{subfigure} \\
        \caption{The $t$-channel gluon exchange contribution.}
        \label{fig:subfig_b}
     \end{subfigure}
        \begin{subfigure}{0.49\linewidth}
            \begin{subfigure}[t]{1.0\textwidth}
            \centering
            \includegraphics[width=1.0\textwidth]{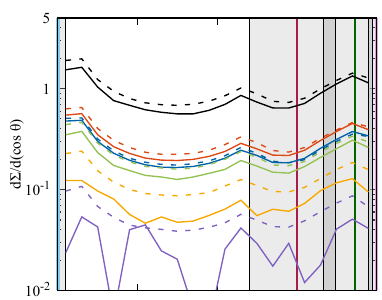}
            \end{subfigure} \\
            \begin{subfigure}[t]{01.\textwidth}
            \centering
            \includegraphics[width=1.0\textwidth]{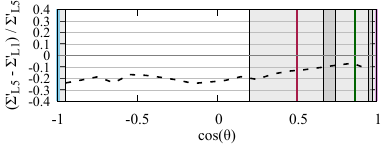}
        \end{subfigure} \\
        \caption{The $s$-channel gluon exchange contribution.}
        \label{fig:subfig_a}
    \end{subfigure}
     \caption{ The differential cross section $\mathrm{d}\Sigma /
       \mathrm{d}(\cos{\theta})$ of the different contributions to the
       $q\bar{q} \to q\bar{q}$ process, in the jets with a recoil
       configuration, and broken down by multiplicity. The solid
       curves are obtained using full colour evolution, and the dashed
       curves use strictly leading colour evolution. For the leading
       colour curves, we start the evolution using the leading-colour
       approximation to the hard-scatter density matrix. The locations
       of the hard jets are marked with vertical lines matching the
       colours used in Fig.~\ref{fig:hatta-diagram}. The shaded
       vertical bars indicate the jet regions. Darker shades indicate
       an overlap of multiple jet-cones.}
     \label{fig:phiAsymmetric-cosTheta}
    \end{figure}

\begin{figure}
        \centering
        \captionsetup[subfigure]{oneside,margin={0.02cm,0cm}}
    \begin{subfigure}{0.49\linewidth}
            \begin{subfigure}[t]{1.\textwidth}
            \centering
            \includegraphics[width=1.0\textwidth]{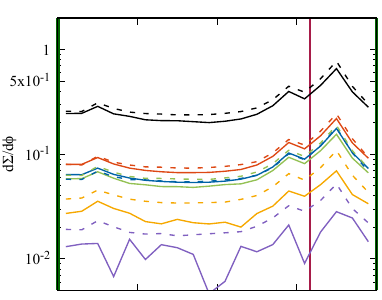}
            \end{subfigure} \\
            \begin{subfigure}[t]{1.\textwidth}
            \centering
            \includegraphics[width=1.0\textwidth]{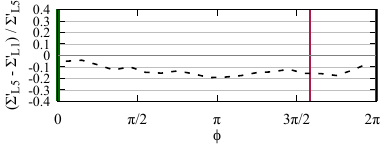}
        \end{subfigure} \\
        \caption{The $t$-channel gluon exchange contribution.}
        \label{fig:subfig_b}
     \end{subfigure}
        \begin{subfigure}{0.49\linewidth}
            \begin{subfigure}[t]{1.0\textwidth}
            \centering
            \includegraphics[width=1.0\textwidth]{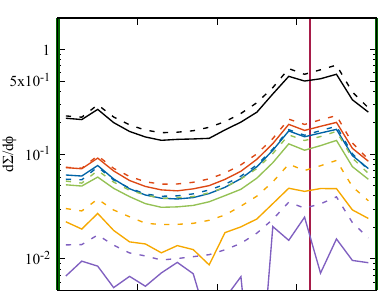}
            \end{subfigure} \\
            \begin{subfigure}[t]{01.\textwidth}
            \centering
            \includegraphics[width=1.0\textwidth]{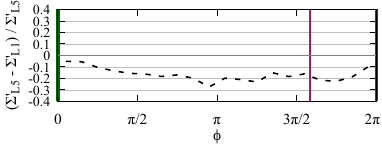}
        \end{subfigure} \\
        \caption{The $s$-channel gluon exchange contribution.}
        \label{fig:subfig_a}
    \end{subfigure}
     \caption{ The differential cross section $\mathrm{d}\Sigma /
       \mathrm{d}\phi$ of the different contributions to the $q\bar{q}
       \to q\bar{q}$ process, in the jets with a recoil configuration,
       and broken down by multiplicity. The solid curves are obtained
       using full colour evolution, and the dashed curves use strictly
       leading colour evolution. For the leading colour curves, we
       start the evolution using the leading-colour approximation to
       the hard-scatter density matrix. The locations of the hard jets
       are marked with vertical lines matching the colours used in
       Fig.~\ref{fig:hatta-diagram}.}
     \label{fig:phiAsymmetric-phi}
    \end{figure}

 We conclude by considering the case where the jets are produced with
 a recoil, such as may occur in vector boson fusion to produce a Higgs
 or a pair of weak bosons at the LHC. In this configuration both final
 state jets point in the general direction of the incoming $q_1$, thus
 not favouring any particular colour configuration. The goal is to
 test whether this will spoil the agreement between full colour and
 leading colour for the radiation patterns of the $t$-channel gluon
 exchange. We find the evolution of the individual $|01\>\<01|$ and
 $|10\>\<10|$ contributions do not show any remarkable features,
 likely due to the angles of the connected pairs being of similar
 sizes, so we omit them for brevity. Instead, we focus on the $s$- and
 $t$- channel gluon exchange contributions, in
 Figs.~\ref{fig:phiAsymmetric-cosTheta}
 and~\ref{fig:phiAsymmetric-phi}.

  In general, we find that the full colour cross section is enhanced by $\sim 10\%$
  relative to the leading colour one in the region near the three
  closer jets $q_1$, $q_3$, and $\bar{q}_4$ for both gluon exchange
  channels. This is clearly visible in
  Fig.~\ref{fig:phiAsymmetric-cosTheta}. We also observe
  significantly different radiation distributions between full and leading
  colour, which are most clearly visible by looking at individual
  multiplicities. For example, in Fig.~\ref{fig:phiAsymmetric-phi},
  $t$-channel gluon exchange, the two-emission orange curves have
  different shapes. There are multiple competing colour configurations
  that radiate in this region, which might explain why full colour
  evolution, which can access all configurations, may enhance
  radiation. This kinematic configuration provides us with a case where the strictly leading colour approximation for hard processes involving $t$-channel gluon exchange fails. 

\section{Conclusions and outlook}
\label{sec:Conclusions}

We have investigated the full colour evolution of jet processes
differentially in the inter-jet radiation. This study provides crucial
insight into how subleading colour effects affect collider observables
beyond the rather inclusive gaps-between-jets measurements, which
sometimes turn out to be sensitive to subleading colour only via an
overall normalization. In fact, we generally find that there are
intricate radiation patterns between jets, and some of the full colour
effects cancel in observables only upon integrating over large enough
patches of solid angle. Our intention is to follow up this study by
looking at more differential observables, including full collider
cross sections involving the entire hard-scatter matrix elements, and
to also assess the impact of Coulomb exchanges, which we are able to
compute using \texttt{CVolver}, as demonstrated in the appendix. At
future lepton colliders multi-jet final states, such as those
originating from hadronic decays of electroweak final states, will be
affected by sub-leading colour, and we have calculated the $1/\Nc^2$
suppressed interference contribution. It will be interesting to
confront this with existing colour reconnection models, in particular
those motivated by soft gluon evolution
\cite{Gieseke:2017clv,Gieseke:2018gff}. Ultimately, it is desirable
that colour reconnection and hadronization models be studied in the
context of full colour evolution \cite{Platzer:2022jny}. For the
future, it remains for us to include hard-collinear physics in
\CVolver, which will allow us to describe collinear sensitive
observables as detailed in \cite{Forshaw:2019ver,Platzer:2022jny}.

\clearpage

\appendix

\section{Coulomb gluon exchange}
\label{app:coulomb}
\begin{figure}[H]
\centering
\begin{subfigure}[t]{0.48\textwidth}
\centering
\includegraphics[width=1.0\textwidth]{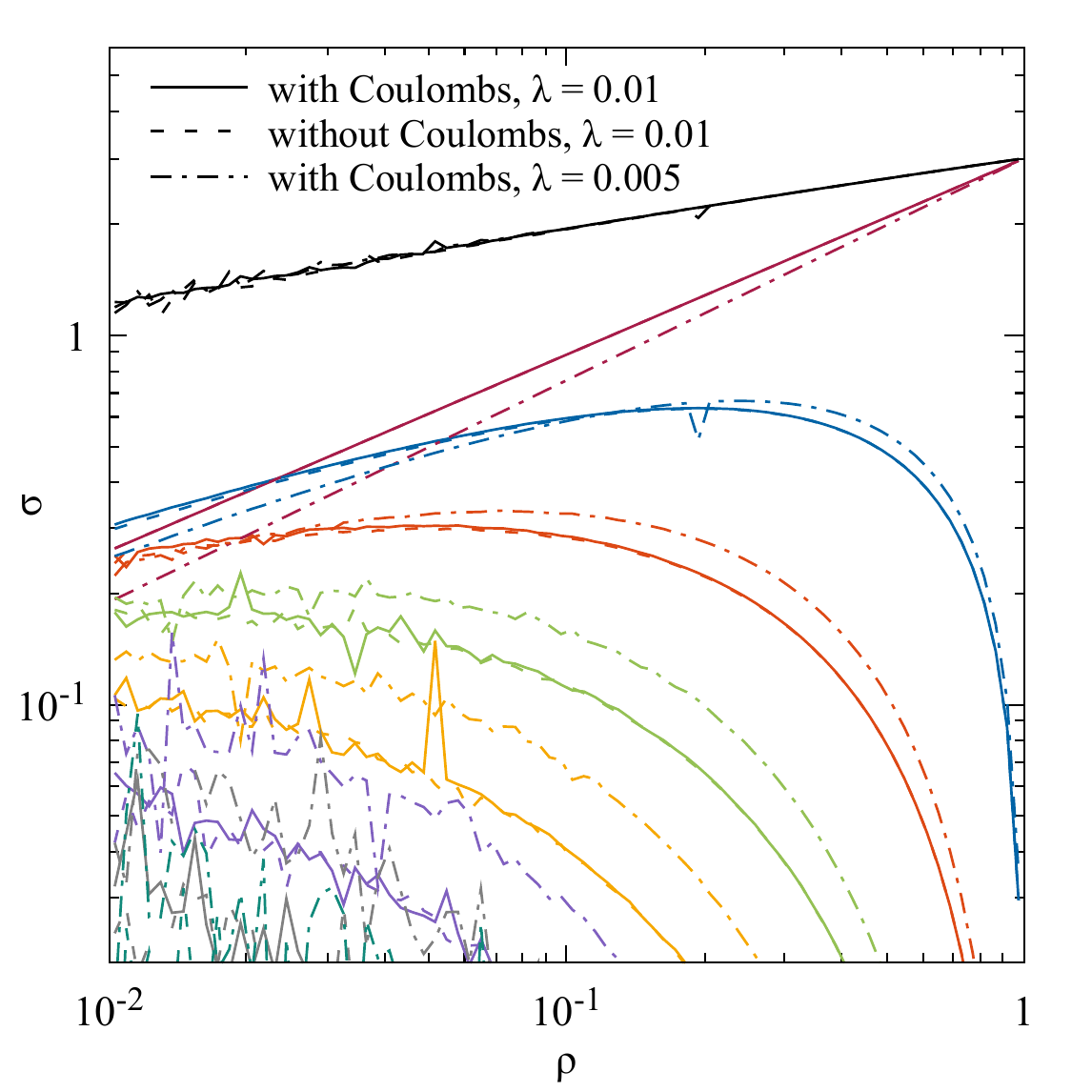}
\caption{$Z \to q \bar{q}$}
\end{subfigure} \hfill
\begin{subfigure}[t]{0.48\textwidth}
\centering
\includegraphics[width=1.0\textwidth]{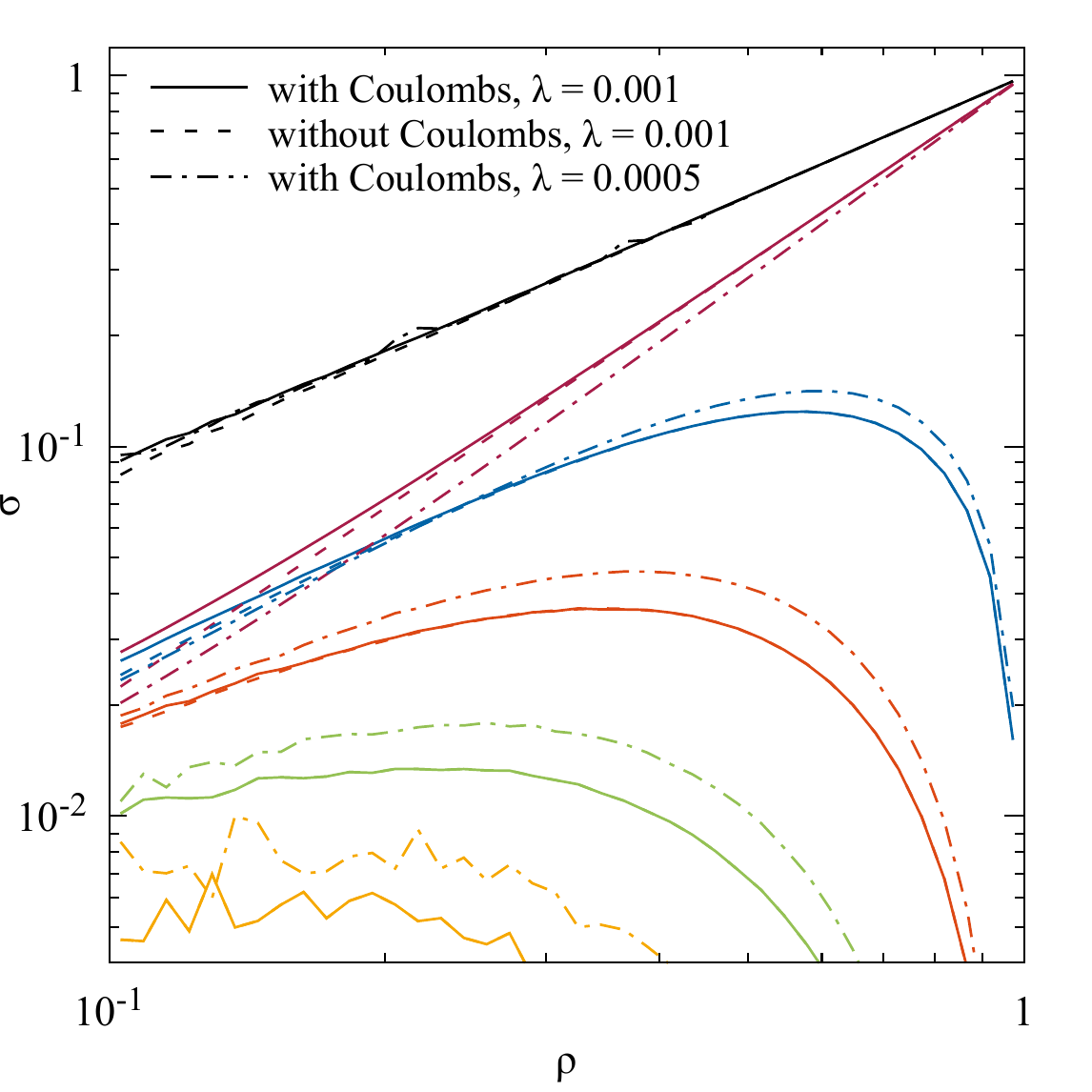}
\caption{$t$-channel gluon exchange $q\bar{q} \to q\bar{q}$, asymmetric configuration}
\end{subfigure}
\caption{Testing the impact of including Coulomb gluons.}
\label{fig:control}
\end{figure}

\begin{figure}[H]
\centering
\includegraphics[width=0.7\textwidth]{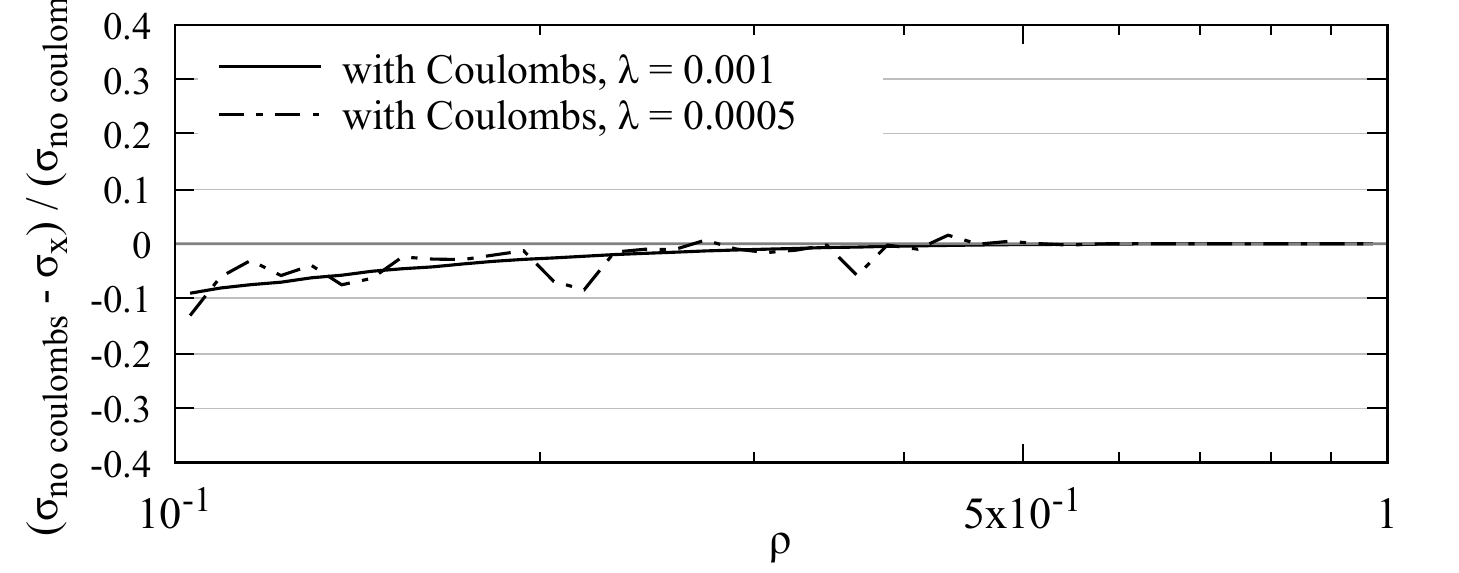}
\caption{Including Coulombs in $q\bar{q} \to q\bar{q}$ induces a $10\%$ effect at $\rho =0.1$}
\label{fig:coulomb-residual}
\end{figure}

There is very considerable interest in exploring the role played by
Coulomb (Glauber) gluons in QCD scattering processes. The discovery of
super-leading logarithms \cite{Forshaw:2006fk,Forshaw:2008cq} and the
related breakdown of soft-collinear factorization
\cite{Catani:2011st,Forshaw:2012bi,Schwartz:2017nmr} has led to
progress in computing and resumming super-leading logarithms
\cite{Keates:2009dn,Becher:2021zkk,Becher:2023mtx,Boer:2024hzh}
including a first demonstration of their numerical importance in
hadron-hadron collisions \cite{Becher:2024nqc}.  At present,
\CVolver\ is able to include Coulomb exchanges in the soft anomalous
dimension, however due to the fact that we do not include
hard-collinear physics the super-leading logarithms will formally
diverge (in fact they grow logarithmically with the collinear
cutoff). In Fig.~\ref{fig:control} we show the effect of turning
Coulomb exchanges on, first in $Z \to q\bar{q}$, which has a
colourless initial state, and also in $q\bar{q} \to q\bar{q}$. In both
cases we consider the back-to-back kinematic configuration considered
in the main text and we show results for different values of the
collinear cutoff. We see clearly the non-trivial cancellation of
Coulomb exchanges in the former and their non-cancellation in the
latter. Fig.~\ref{fig:coulomb-residual} is the residual plot $q\bar{q}
\to q\bar{q}$ and we can see that the result is collinear cutoff
independent at these relatively high values of $\rho$. We do not
expect this to persist to low values of $\rho$.

\acknowledgments This work has received funding from the U.K. Science
and Technology Facilities Council grant no. ST/X00077X/1. FTG is
supported by the Royal Society through Grant URF/R1/201500. We thank
Jack Holguin for fruitful discussions, and Matthew De Angelis for
earlier contributions to the program. The numerical results presented
in this paper have been obtained on the computing clusters of the
Particle Physics Groups of Universit\"at Wien and The University of
Manchester. We are grateful for having been able to use these
facilities. FTG acknowledges the kind hospitality of the Theoretical
Physics Group of the Institute of Physics of the Universit\"at Graz.

\bibliography{refs}

\end{document}